\documentclass[aps,prb,twocolumn,footinbib,showpacs,superscriptaddress,preprintnumbers,amsmath,amssymb]{revtex4-1}

\usepackage{graphicx}
\usepackage{subfigure}
\usepackage{dcolumn}
\usepackage{bm}
\usepackage{verbatim} 
\usepackage{color}
\usepackage{float}

\begin{document}
\title{Spin-resolved electron waiting times in a quantum dot spin valve}

\author{Gaomin Tang}
\affiliation{Department of Physics and the Center of Theoretical and Computational Physics, The University of Hong Kong, Hong Kong, China}
\author{Fuming Xu}
 \email{xufuming@szu.edu.cn}
\affiliation{Shenzhen Key Laboratory of Advanced Thin Films and Applications, College of Physics and Energy, Shenzhen University, Shenzhen 518060, China}
\author{Shuo Mi}
\affiliation{Department of Applied Physics, Aalto University, 00076 Aalto, Finland}
\affiliation{Department of Physics and the Center of Theoretical and Computational Physics, The University of Hong Kong, Hong Kong, China}
\author{Jian Wang}
 \email{jianwang@hku.hk}
\affiliation{Department of Physics and the Center of Theoretical and Computational Physics, The University of Hong Kong, Hong Kong, China}

\begin{abstract}
We study the electronic waiting time distributions (WTDs) in a non-interacting quantum dot spin valve by varying spin polarization and the noncollinear angle between the magnetizations of the leads using scattering matrix approach. Since the quantum dot spin valve involves two channels (spin up and down) in both the incoming and outgoing channels, we study three different kinds of WTDs, which are two-channel WTD, spin-resolved single-channel WTD and cross-channel WTD. We analyze the behaviors of WTDs in short times, correlated with the current behaviors for different spin polarizations and noncollinear angles. 	
	Cross-channel WTD reflects the correlation between two spin channels and can be used to characterize the spin transfer torque process. We study the influence of the earlier detection on the subsequent detection from the perspective of cross-channel WTD, and define the influence degree quantity as the cumulative absolute difference between cross-channel WTDs and first passage time distributions to quantitatively characterize the spin flip process.
	The influence degree shows a similar behavior with spin transfer torque and can be a new pathway to characterize spin correlation in spintronics system. 
\end{abstract}

\pacs{73.23.-b,	
73.63.-b,	
72.70.+m,	
73.23.Hk,	
72.25.-b,	
85.75.-d	
}

\maketitle

\section{introduction}
Spintronics, which utilizes the spin degree of freedom to process and store information in nanostructured devices, has received intensive research in the past decades \cite{spin1, spin2, spin3}. In spintronics, the magnetic tunnel junction (MTJ) is of particular interest, which typically consists of two ferromagnetic leads separated by an insulating layer such as Fe/MgO/Fe junction \cite{MTJ1, MTJ2, KGong}. The spin-polarized current varies with the spin polarization and the relative directions of magnetization in the magnetic layers. In general, the tunneling current of parallel configuration of the two magnetic layers is much larger than that of antiparallel configuration, and this is the so-called tunnel magnetoresistance (TMR) \cite{TMR1, TMR2, TMR3, TMR4, TMR5, KGong}. As predicted independently by Slonczewski \cite{STT1} and Berger \cite{STT2} in 1996, spin current is not conserved through the MTJ with noncollinear magnetizations in the magnetic layers, which can induce a spin transfer torque (STT) on the magnetization \cite{Theodonis1, Theodonis2, Theodonis3, STT11, STT22}. STT has been applied on spintronic devices such as STT-MRAM, which employs the STT instead of the magnetic field to control the magnetization and hence has lower power consumption \cite{STT3}. Other investigations of MTJ include spin dependent Seebeck effect in the thermoelectric engine \cite{DiVentra, Barnas, Bauer2, Bauer3, gm5}, angle dependent conductance \cite{angle1, angle2, angle3}, adiabatic pumping \cite{pump, FCS-STT}, etc.. The quantum dot (QD) spin valve is related to MTJ, and has both the TMR and STT effect as well. If one tunes the QD levels far away from the resonant condition, it can mimic the behaviors of MTJ with insulating scattering region. Yu {\it et. al.} has shown that the off-resonant behaviors of the spin torque of QD spin valve is the same as that of MTJ \cite{YunjinYu}.

	Current and its fluctuation are typical characterizations of quantum transport properties in nano devices \cite{Blanter}. A more general description beyond current and fluctuation should resort to the formalism of full-counting statistics (FCS), which can give a full scenery of probability distribution of transferred charges and all zero-frequency cumulants at long times \cite{Levitov1, Levitov2, Flindt1, wavepacket, Flindt2, Fernando, Flindt3, gm2, JS3, Ruben1, gm3, gm4, gm6, Ruben2, Michael}. FCS of charge and STT in MTJ and QD spin valve has received intensive attentions \cite{FCS-STT, gm2, FCS-MTJ}, and magnetization switching probability can also be evaluated via FCS \cite{switch}. With the rapid development of single-electron devices \cite{SET}, deeper understanding of important information on short time physics becomes possible. However, FCS usually deals with collective behaviors of many electrons at long times and the short-time particle dynamics is lost.
	As a complement to FCS, electronic WTD has been developed to characterize the short-time correlation in mesoscopic conductors, which is the probability density of delay times between two subsequent charge transfers \cite{kampen}. WTDs have been studied for systems governed by either Markovian \cite{WTD_Brandes, WTD_Yan, WTD2011, WTD_rajabi, WTD_spin_valve, WTD_Kosov, WTD_turnstile, WTD_CPS} or non-Markovian \cite{WTD_non-Markovian} master equations. The scattering matrix formalism \cite{wavepacket} has been developed to calculate WTDs under both constant voltage \cite{WTD2012, WTD_TB} and periodic drive \cite{WTD_Floquet, WTD_Leviton, WTD2}. 
A quantum theory of waiting time clock has been developed in order to measure WTDs experimentally \cite{WTD_clock}. Generalization to multiple channels \cite{WTD2014, WTD_correlated} has been made, and the formalism of joint WTD \cite{WTD_correlated} which characterizes the correlation between subsequent times has been established. Spin-averaged WTD in a QD spin valve has been studied by B. Sothmann \cite{WTD_spin_valve}.

Since spintronic phenomenon plays an indispensable role in fundamental research and industrial application, investigation of WTD in spintronic system is very important.
	Spin-resolved WTD of spintronic system, which involves at least two channels (spin up and down), is still lack of study and the lacuna should be filled.
We note that QD spin valve or MTJ are earlier examples in the family of spintronics. In this work, we employ multi-channel WTD formalism and study WTDs and cross-channel WTDs of the QD spin valve using scattering matrix approach. The scattering matrix approach requires the electronic reservoir to have a linear dispersion with respect to the momentum in the transport window and the system at zero temperature. We employ the nonequilibrium Green's function technique, which does not rely on week coupling between QD and electrodes, to get the scattering amplitude. The behaviors of two channel, spin-$\sigma$ and cross-channel WTDs are numerically calculated with respect to noncollinear  angle and spin polarization, and their behaviors at initial short times are identified and explained. 
The difference between cross-channel WTD and corresponding first passage time distribution (FPTD) reveals the influence of the first detection on the subsequent one, and indicates the correlation between spin channels. In order to characterize the correlation strength between spin channels quantitatively, we introduce the 'influence degree' quantity as the cumulative absolute difference between cross-channel WTDs and FPTDs. We find that the influence degree vanishes for collinear configurations, and reaches its maximum near noncollinear angle $\theta =\pi/2$ in which STT also achieves its maximal value. Since spin correlation strength increases with increasing spin polarization, influence degree is an increasing function with respect to the spin polarization.

	The paper is organized as follows. In Sec.\ref{sec2}, the system setup and theoretical formalism of two-channel WTD are introduced. We also present spin-resolved waiting time clock in this section. In Sec.\ref{sec3}, we show the numerical results of WTD  by varying the spin polarization and the angle between the magnetizations of the leads in detail, accompanied with discussion and analysis. We finally summarize our work in Sec.~\ref{sec4}.

\section{model and theoretical formalism}\label{sec2}
\subsection{Magnetic tunnel junction}

\begin{figure}
  \includegraphics[width=2.6in]{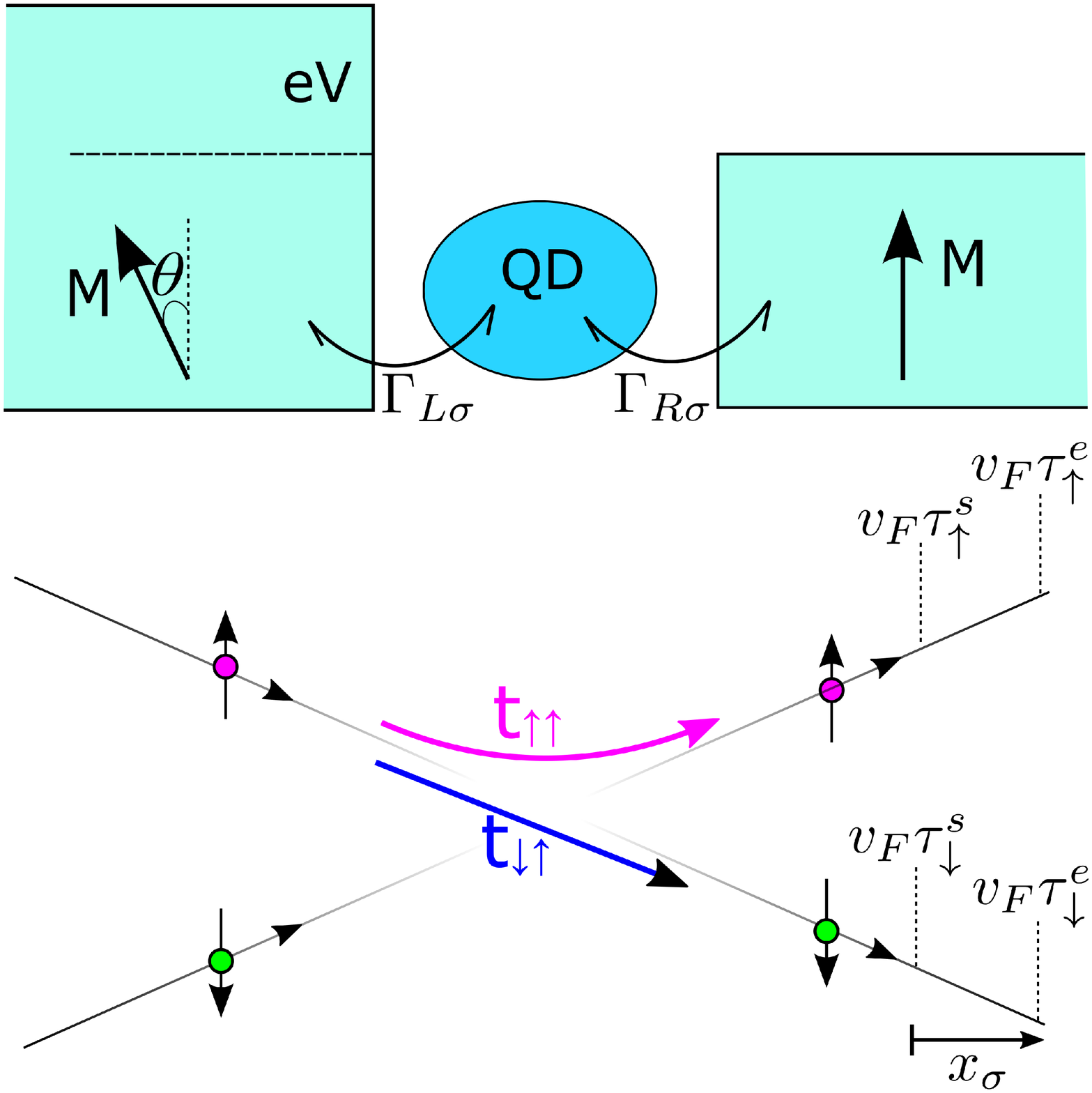} \\
  \caption{(Color online) Schematic illustration of a spin valve in which a QD coupled to its left and right ferromagnetic leads $\alpha = L,R$ through coupling strengths $\Gamma_{\alpha\sigma}$. The magnetization of the right lead is along the $z$-axis, while that of the left lead is along the $z'$-axis at a noncollinear angle of $\theta$ to the z-axis.  The transmitted electrons with spin $\sigma$ are detected in the outgoing channels (right lead) at different positions $x_\sigma \in [v_F \tau_\sigma^s, v_F \tau_\sigma^e]$. }
  \label{fig1}
\end{figure}

The spin valve we consider consists of a QD coupled to its left and right ferromagnetic leads $\alpha = L,R$, with the magnetization of the left lead at a noncollinear angle of $\theta$ to the magnetization of the right lead. We consider a large QD so that the Coulomb interaction effect can be neglected. The system Hamlitonian reads as
\begin{equation}
H= H_S + \sum_{\alpha=L,R} (H_\alpha + H_{\alpha S}) .
\end{equation}
Here, the Hamiltonian of the non-interacting QD is expressed as
\begin{equation}
H_S = \sum_{\sigma} \epsilon_{\sigma} d_{\sigma}^\dag d_{\sigma}, 
\end{equation}
where $\epsilon_{\uparrow}$ and $\epsilon_{\downarrow}$ can be different for a quantum spin Hall (QSH) QD \cite{QSH_QD1, QSH_QD2, QSH_QD3} which will be elaborated in discussing the waiting time clock.  
And $H_\alpha$ describes the Hamiltonians of the left and right leads in the local reference frame with the form, 
\begin{equation}
H_\alpha = \sum_{k\sigma} \epsilon_{k\alpha\sigma} c_{k\alpha\sigma}^\dag c_{k\alpha\sigma},
\end{equation}
where $\epsilon_{k\alpha\sigma}$ is the energy of an electron with spin $\sigma$ and wave number $k$ in the $\alpha$ ferromagnetic lead. 
The coupling Hamiltonians between QD and the left and right leads are \cite{angle3}
\begin{align}
H_{LC}& =\sum_k c_{kL}^\dag t_L {\cal R} d +\rm{H.c.} , \notag \\
H_{RC}& =\sum_k c_{kR}^\dag t_R d +\rm{H.c.} ,
\end{align}
respectively, where we used the abbreviations 
$c_{k\alpha}^\dag=(c_{k\alpha\uparrow}^\dag, c_{k\alpha\downarrow}^\dag)$, and $d^\dag=(d_{\uparrow}^\dag,  d_{\downarrow}^\dag)$. Here, $t_{\alpha}={\rm diag}(t_{\alpha\uparrow}, t_{\alpha\downarrow})$ is the hopping matrix elements between the QD and the spin $\sigma$ electronic states in the lead $\alpha$ when $\theta=0$. The rotation matrix ${\cal R}$ from the Bogoliubov transformation is applied to diagonalize the Hamiltonian of the left lead and has the form \cite{gm2}
\begin{equation}  \label{rotation}
{\cal R}=\begin{pmatrix}
\cos\frac{\theta}{2} & -\sin\frac{\theta}{2} \\
\sin\frac{\theta}{2} & \cos\frac{\theta}{2}
\end{pmatrix} .
\end{equation}
The coupling strength between the QD and leads in the collinear configuration is described by $\Gamma_{\alpha\sigma} =2\pi |t_{\alpha}|^2\rho_{\alpha\sigma}$, and we set $\Gamma_\alpha = (\Gamma_{\alpha\uparrow}+\Gamma_{\alpha\downarrow})/2$. $\rho_{\alpha\sigma}$ is the density of states for the spin $\sigma$ electrons in the lead $\alpha$. The spin polarization $p_\alpha$ in the lead $\alpha$ is given by,
\begin{equation}
p_\alpha= \frac{\rho_{\alpha\uparrow}-\rho_{\alpha\downarrow}}{\rho_{\alpha\uparrow}+\rho_{\alpha\downarrow}} = \frac{\Gamma_{\alpha\uparrow}-\Gamma_{\alpha\downarrow}}{\Gamma_{\alpha\uparrow}+\Gamma_{\alpha\downarrow}}.
\end{equation}
$p_\alpha= 0$ indicates that lead $\alpha$ is a normal metal, and $p_\alpha= 1$ denotes a half-metallic ferromagnet.
Then the coupling strength can be written as $\Gamma_{\alpha\sigma}=\Gamma_\alpha (1+\sigma p_\alpha)/2$, with $\sigma=1$ for spin-up and $\sigma=-1$ for spin-down.
In this work, we assume the system is symmetric and both leads have the same coupling strength, $\Gamma_L=\Gamma_R\equiv \Gamma$, and the same polarization, $p_L=p_R \equiv p$.

	The retarded Green's function of the central QD in spin space is
\begin{equation}
{\bf G}^r(E) = (E-H_0-{\cal R} \Sigma_L^r {\cal R}^\dag -\Sigma_R^r)^{-1},
\end{equation}
where $H_0={\rm diag}(\epsilon_\uparrow, \epsilon_\downarrow)$, and the retarded self-energy in the lead $\alpha$ is $\Sigma_{\alpha\sigma}^r=-i\Gamma_{\alpha\sigma}/2$.
The transmission matrix from the left lead to the right lead is ${\bf T}={\bf G}^r {\cal R} {\bf \Gamma}_L {\cal R}^\dag {\bf G}^a {\bf \Gamma}_R$, with ${\bf \Gamma}_\alpha = {\rm diag}(\Gamma_{\alpha\uparrow}, \Gamma_{\alpha\downarrow})$, and ${\bf G}^a = [{\bf G}^r]^\dag$.
The transmission amplitude matrix, which consists of the scattering matrix elements relating the left and right leads, can be obtained using the Fisher-Lee relation and expressed as  \cite{Datta,JianWang,off-diagonal},
\begin{equation}
{\bf t} = \begin{pmatrix}
t_{\uparrow\uparrow} & t_{\uparrow\downarrow} \\
t_{\downarrow\uparrow} & t_{\downarrow\downarrow}
\end{pmatrix} 
= \sqrt{{\bf \Gamma}_R} {\bf G}^r {\cal R} \sqrt{{\bf \Gamma}_L},
\end{equation}
with its component $t_{\sigma\sigma'}$ denoting the transmission amplitude from spin $\sigma'$ in the left lead to spin $\sigma$ in the right lead. The explicit energy $E$ dependence of the transmission amplitude matrix is suppressed for notational simplicity. The QD spin valve is driven out of equilibrium by applying a constant voltage bias $V$. The transport window is $[E_F, E_F+eV]$ with $E_F$ the Fermi level at zero temperature. 
The spin current in the left and right leads are, respectively, expressed as,
\begin{equation}
I_{L\sigma} = \int_{0}^{eV} \left[{\bf t}^\dag {\bf t} \right]_{\sigma\sigma} dE , \quad
I_{R\sigma} = \int_{0}^{eV} \left[{\bf t} {\bf t}^\dag \right]_{\sigma\sigma} dE .
\end{equation}
The particle current through the system is expresses as
\begin{equation}
I = \int_{0}^{eV} {\rm Tr}\left[{\bf t} {\bf t}^\dag \right] dE .
\end{equation}
The spin transfer torque has the expression as, \cite{YunjinYu}
\begin{equation}
{\rm STT} = \int_{0}^{eV} {\rm Tr}\left[ {\bf G}^r (i\Sigma_L^a {\cal \bar{R}}- i{\cal \bar{R}}\Sigma_L^r ) {\bf G}^a  {\bf \Gamma}_R \right] dE .
\end{equation}
with 
\begin{equation}  \label{rotation}
{\cal \bar{R}}=\begin{pmatrix}
-\sin\theta  & \cos\theta  \\ \cos\theta & \sin\theta
\end{pmatrix} .
\end{equation}

\subsection{Waiting time distributions}
In this subsection, we discuss the formalism to calculate waiting times between successive electrons detected in the right lead. The system has two incoming and two outgoing channels, namely, spin up and spin down. If one detects an electron at a starting time $\tau^s$, the conditional probability density of detecting the successive electron at an ending time $\tau^e$ is the two-channel WTD ${\cal W}(\tau^s, \tau^e)$. The detection involved in the two-channel WTD does not differentiate the electron spin. One can also define the spin-resolved WTD ${\cal W}_{\sigma\sigma'}(\tau^s, \tau^e)$, which is the conditional probability density to detect a spin $\sigma'$ electron at an ending time $\tau^e$ on the condition that the starting detection of spin $\sigma$ electron occurred at the earlier time $\tau^s$. If the two successively detected electrons possess the same spin, it is the spin-resolved single-channel WTD, while if the two successive electrons have different spins, one can define it as the cross-channel WTD \cite{WTD_correlated}. Since the dc case is considered here, WTD only depends on the time difference $\tau = \tau^e-\tau^s$ due to the time translational symmetry, and one can write the above defined WTDs as ${\cal W}(\tau)$ and ${\cal W}_{\sigma\sigma'}(\tau)$, respectively.
Before coming back to the discussion of WTDs, we first discuss the idle time probability (ITP) which plays the role of the generating function of WTDs.
		
	We use the scattering matrix approach, which was initially developed by F. Hassler {\it et al.} \cite{wavepacket} and then generalized to multi-channel case by D. Dasenbrook {\it et al.} \cite{WTD_correlated}, to evaluate the ITPs in non-interacting systems at zero temperature. The scattering matrix approach requires a linear dispersion relation with respect to the momentum in the transport window $[E_F, E_F+eV]$,
\begin{equation}
E(k) = \hbar k v_F ,
\end{equation}
where the energy $E(k)$ is measured with respect to the Fermi level and $v_F$ is the Fermi velocity. We assume that the Fermi velocities for spin up and down electrons are the same and no spin bias is present in this work. 
Instead of considering the probability of no spin $\sigma$ electrons detected in the time intervals $[\tau_\sigma^s, \tau_\sigma^e]$, one can consider the probability of detecting no electrons in the spatial interval $[v_F\tau_\sigma^s, v_F\tau_\sigma^e]$. 
We define the single-particle projection operator
\begin{equation}
\widehat{{\cal Q}}_\sigma = \int_{v_F\tau_\sigma^s}^{v_F\tau_\sigma^e} \hat{b}_\sigma^\dag (x) \hat{b}_\sigma(x) dx ,
\end{equation}
which measures the probability of finding a spin $\sigma$ electron in the spatial interval $x_\sigma \in [v_F \tau_\sigma^s, v_F \tau_\sigma^e]$ in the right lead, where $\hat{b}_\sigma^{(\dag)} (x)$  annihilate (create) spin $\sigma$ electrons at position $x$.
The generalized ITP \cite{WTD_correlated}, $\Pi(\tau_\uparrow^s, \tau_\uparrow^e; \tau_\downarrow^s, \tau_\downarrow^e)$, is the joint probability that no spin $\sigma$ electrons are detected during the time intervals $[\tau_\sigma^s, \tau_\sigma^e]$.
It can be expressed as the expectation value of the normal-ordered exponent of $-\sum_\sigma \widehat{{\cal Q}}_\sigma$,
\begin{equation}
\Pi(\tau_\uparrow^s, \tau_\uparrow^e; \tau_\downarrow^s, \tau_\downarrow^e) = \langle : e^{-\sum_\sigma \widehat{{\cal Q}}_\sigma} : \rangle ,
\end{equation}
with $:\cdots:$ denoting the normal-ordering of operators \cite{WTD_correlated}. One may evaluate the average and obtain the ITP in a determinant form \cite{WTD_correlated},
\begin{equation} \label{ITP}
\Pi(\tau_\uparrow^s, \tau_\uparrow^e; \tau_\downarrow^s, \tau_\downarrow^e) = \det\left( {\bf I} -{\bf Q}_{(\tau_\sigma^s, \tau_\sigma^e)} \right) .
\end{equation}
The matrix ${\bf Q}$ is a $2$-by-$2$ block matrix in the spin space with the form \cite{WTD_correlated},
\begin{equation}
{\bf Q}_{(\tau_\sigma^s, \tau_\sigma^e)}(E,E') = {\bf t}^\dag (E) {\bf K}(E-E'){\bf t}(E').
\end{equation}
Here, the kernel matrix is diagonal in spin space and reads as \cite{WTD_correlated} 
\begin{equation}
{\bf K}_{\sigma\sigma}(E) = \frac{\kappa}{\pi} e^{-iE(\tau_\sigma^s + \tau_\sigma^e)/2} \frac{\sin[E(\tau_\sigma^s- \tau_\sigma^e)/2]}{E} .
\end{equation}
In calculating the determinant, Eq.~\eqref{ITP}, we have divided the transport window into $N$ energy elements, each with size $\kappa = eV/N$. A large $N$ should be taken to ensure the numerical convergence.

	The two-channel ITP $\Pi(\tau^s,\tau^e)$, the probability of detecting no electron regardless of the spin degree in any of the outgoing channels during a time interval $[\tau^s,\tau^e]$, can be obtained from the generalized ITP by setting $\tau_\uparrow^e=\tau_\downarrow^e=\tau^e$, $\tau_\uparrow^s=\tau_\downarrow^s=\tau^s$. The ITP for a single spin $\sigma$ channel can be obtained from the generalized ITP as $\Pi_\sigma(\tau_\sigma^s, \tau_\sigma^e)\equiv \Pi(\tau_\sigma^s, \tau_\sigma^e; \tau_{\bar{\sigma}}^s=\tau_{\bar{\sigma}}^e)$. Here and below, we use notation $\bar{\sigma}$ to denote the spin index which is different from $\sigma$, with $\bar{\sigma}=\downarrow$ for $\sigma=\uparrow$, and $\bar{\sigma}=\uparrow$ for $\sigma=\downarrow$.

	The joint probability density of detecting two successive electrons both at $\tau^s$ and $\tau^e$ is equal to the WTDs multiplied by the probability density of a detection event at $\tau^s$. For the uni-directional quantum transport considered in this work, the probability density of a detection at $\tau^s$ is simply the electronic current $I(\tau^s)$ without distinguishing spin or spin current $I_{R\sigma}(\tau^s)$ for a specific spin channel. The joint probability density can also be obtained by differentiating the ITPs with respect to both the starting time $\tau^s$ and the ending time $\tau^e$. Then we can get the equations for two-channel WTD, spin-resolved single-channel WTD, and cross-channel WTD, respectively, as \cite{WTD_correlated}
\begin{align}
I(\tau^s) {\cal W}(\tau^s,\tau^e) &= - \partial_{\tau^s}\partial_{\tau^e} \Pi(\tau^s, \tau^e) , \\
I_{R\sigma}(\tau^s) {\cal W}_{\sigma\sigma}(\tau^s, \tau^e) &= -\partial_{\tau^s}\partial_{\tau^e} \Pi_\sigma(\tau^s, \tau^e) , \\
I_{R\sigma}(\tau^s) {\cal W}_{\sigma\bar{\sigma}}(\tau^s, \tau^e) &= -\partial_{\tau^s}\partial_{\tau^e} \Pi(\tau_\sigma^s,\tau^e;\tau^s,\tau^e)\big|_{\tau_\sigma^s=\tau^s} .
\end{align}
For the dc transport at zero temperature, the electronic current is the inverse mean waiting time \cite{WTD2012, WTD_correlated}, so that we have $I(\tau^s) = 1/\langle \tau \rangle$, and $I_{R\sigma}(\tau^s) = 1/\langle \tau_\sigma \rangle$, where $\langle \tau \rangle$ is the average two-channel waiting time, and $\langle \tau_\sigma \rangle$ is the average spin-resolved single-channel waiting time.
	Since the dc quantum transport possesses the time translational symmetry, WTDs and ITPs only depend on the time difference $\tau = \tau^e-\tau^s$, and one can write the above expressions of WTD as \cite{WTD_correlated},
\begin{align}
{\cal W}(\tau) &= \langle \tau \rangle \frac{\partial^2 \Pi(\tau)}{\partial\tau^2}, \\
{\cal W}_{\sigma\sigma}(\tau) &= \langle \tau_\sigma \rangle \frac{\partial^2 \Pi_\sigma(\tau)}{\partial\tau^2},  \\
{\cal W}_{\sigma\bar{\sigma}}(\tau) &= \langle \tau_\sigma \rangle \frac{\partial^2 \Pi(\tau_\sigma^s,\tau^e;\tau^s,\tau^e)}{\partial\tau_\sigma^s \partial\tau^e} \bigg|_{\tau_\sigma^s=\tau^s; \ \tau^e-\tau^s=\tau} .
\end{align}
The formalism presented is used to calculate WTDs for the non-interacting systems at zero temperature and assumes a linear dispersion relation with respect to the momentum in the electronic reservoir. It assumes neither Markovian approximation nor any renewal properties.

	The first passage time distribution (FPTD) ${\cal F}_\sigma(\tau_\sigma^s, \tau')$ is the probability density for the event to occur at a time $\tau'$, in spite of the observation result of the previous time $\tau_\sigma^s$ \cite{kampen, WTD_correlated, gm1}. One can relate the FPTD of the spin $\sigma$ channel with the corresponding ITP through the relation, 
\begin{equation}
1-\int_{\tau_\sigma^s}^{\tau_\sigma^e} {\cal F}_\sigma(\tau_\sigma^s, \tau') d\tau' = \Pi_\sigma(\tau_\sigma^s, \tau_\sigma^e; \tau_{\bar{\sigma}}^s=\tau_{\bar{\sigma}}^e) .
\end{equation}
The time integral in the above equation represents the probability to detect spin $\sigma$ electrons during the time interval $[\tau_\sigma^s, \tau_\sigma^e]$. For dc quantum transport, the FPTD is expressed as \cite{gm1}, 
\begin{equation}
{\cal F}_\sigma(\tau) = -\partial_{\tau} \Pi_\sigma(\tau).
\end{equation}
If the outgoing spin up and down channels are uncorrelated, the detection result of a later time in one channel doesn't depend on the earlier detection in the other channel, so that the cross-channel WTD for uncorrelated spin channels is equal to the FPTD \cite{WTD_correlated}, 
\begin{equation}
{\cal W}_{\sigma\bar{\sigma}}^{\rm uc}(\tau) = {\cal F}_{\bar{\sigma}}(\tau) .
\end{equation}

\begin{figure}
  \includegraphics[width=2.6in]{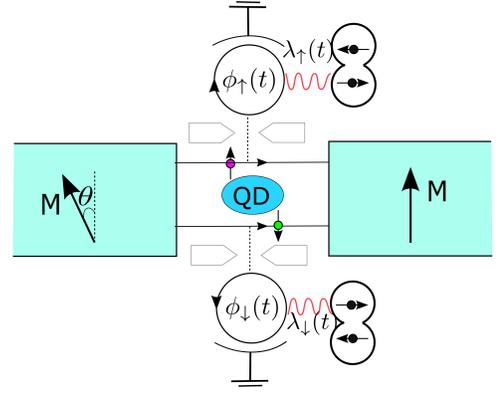} \\
  \caption{(Color online) Schematic plot of spin-resolved waiting time clock. A quantum spin Hall QD is embedded between two ferromagnetic electrodes. Spin-$\sigma$ (spin-up in the plot) electrons can tunnel into the capacitor through a quantum point contact and then interacts with a two-level system. Spin-$\sigma$ electron waiting times can be obtained from monitoring the two-level system by changing interaction strength $\lambda_\sigma(t)$. }
  \label{fig2-clock}
\end{figure}

In order to measure WTD above the Fermi sea experimentally, a quantum formalism of a detector, which is called waiting time clock, has been proposed \cite{WTD_clock}. 
The waiting time clock consists of a mesoscopic capacitor being coupled to a quantum two-level system. The electrons from the system transmit to a chiral edge state in the quantum Hall regime and then tunnel into the capacitor through a quantum point contact. The quantum point contact only transmits the electrons above the Fermi sea. Electrons inside the capacitor interact with a two-level system of which we monitor the coherent precession, and then leave the capacitor. The coupling strength $\lambda(t)$ between the two-level system and the capacitor is tunable and time-dependent. The moment generating function could be obtained from reading the off-diagonal element of the density matrix of the two-level system for different coupling strengths $\lambda$. Then one can get the ITP from the moment generating function, and hence WTD. The chiral edge state is needed here so that the electron can tunnel into the capacitor through a quantum point contact. For the system presented in our work, we consider the QD to be a quantum spin Hall (QSH) quantum dot \cite{QSH_QD1, QSH_QD2, QSH_QD3} (see Fig.~\ref{fig2-clock}) in order to have edge states in the QD spin valve. 
When the the Fermi wavelength are longer than the distance between two ferromagnetic electrodes, the spin-dependent scattering can open a gap to form a dot in a systems such as the double HgTe/CdTe quantum well \cite{double-QWs}. 
One can also use QSH edges in contact to the QD \cite{QSH_QD4, QSH_QD5} to form chiral edge states in the central scattering region. When the Fermi energy of QD is inside the energy gap, electrons only tunnel through the unidirectional spin locked edge state, and one can use one edge to transmit spin up electrons and the other edge to transmit spin down electrons. Then the spin-$\sigma$ WTD can be measured by measuring the two-level system with which the spin-$\sigma$ electrons interact. Waiting time clock involving cross-channel detection is also worth future investigation.

\section{numerical results and discussion} \label{sec3}


\begin{figure}
\centering
  \includegraphics[width=3.4in]{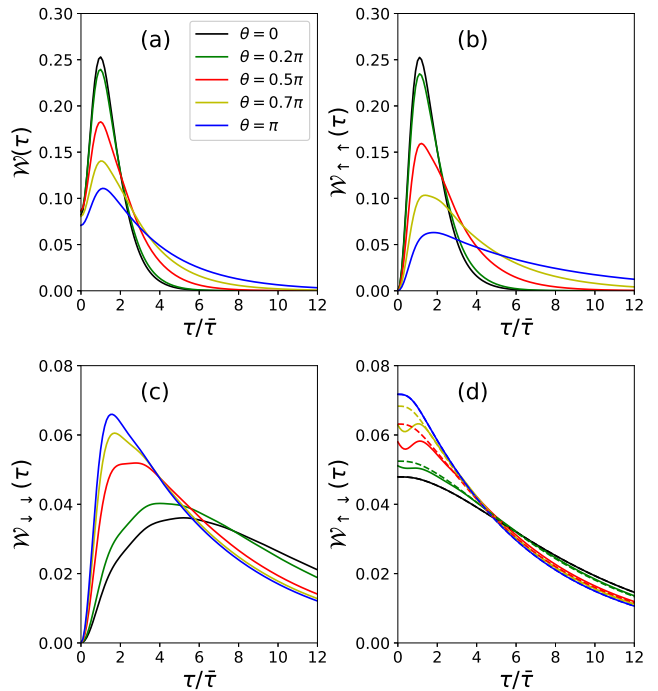} \\
  \caption{(Color online) Two-channel WTD ${\cal W}(\tau)$ [panel (a)], spin-up WTD ${\cal W}_{\uparrow\uparrow}(\tau)$ [panel (b)], spin-down WTD ${\cal W}_{\downarrow\downarrow}(\tau)$ [panel (c)], and cross-channel WTD ${\cal W}_{\uparrow\downarrow}(\tau)$ [panel (d)] are plotted by varying noncollinear angle $\theta$ with spin polarization $p=0.8$. 
The corresponding FPTD for spin down ${\cal F}_{\downarrow}(\tau)$ is plotted with dashed line in panel (d). The waiting time is in units of time $\bar{\tau}=h/(eV)$. }
  \label{fig3}
\end{figure}

\begin{figure}
\centering
  \includegraphics[width=3.4in]{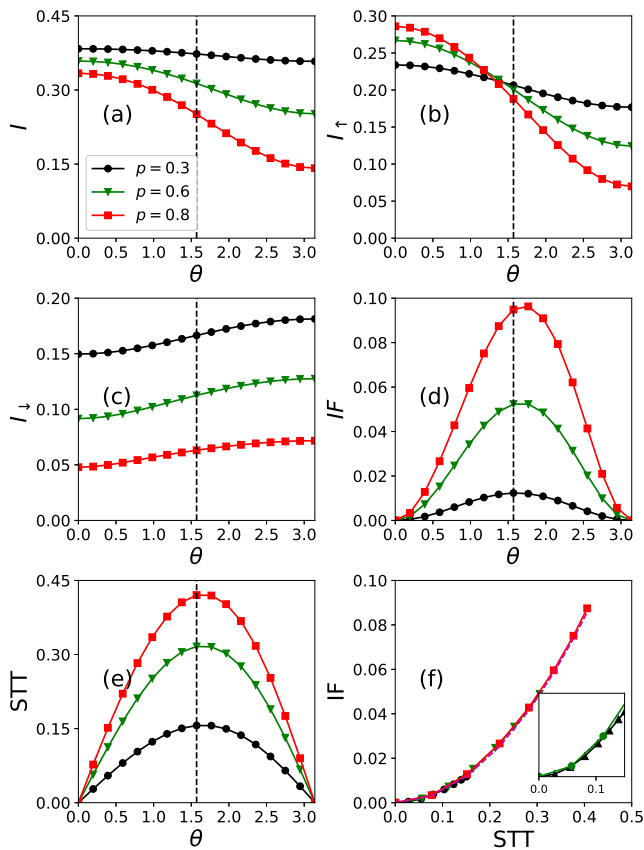} \\
  \caption{(Color online) Charge current [panel (a)], spin up current $I_{R\uparrow}$ [panel (b)], spin down current $I_{R\downarrow}$ [panel (c)], influence degree [panel (d)], and STT [panel (e)] versus noncollinear angle $\theta$ with different spin polarization $p$ are plotted. Influence degree versus STT are plotted in panel (f) with the quadratic fitting ${\rm IF}=0.51*({\rm STT})^2$ shown in dashed lines. }
  \label{fig4}
\end{figure}

\begin{figure}
\centering
  \includegraphics[width=3.4in]{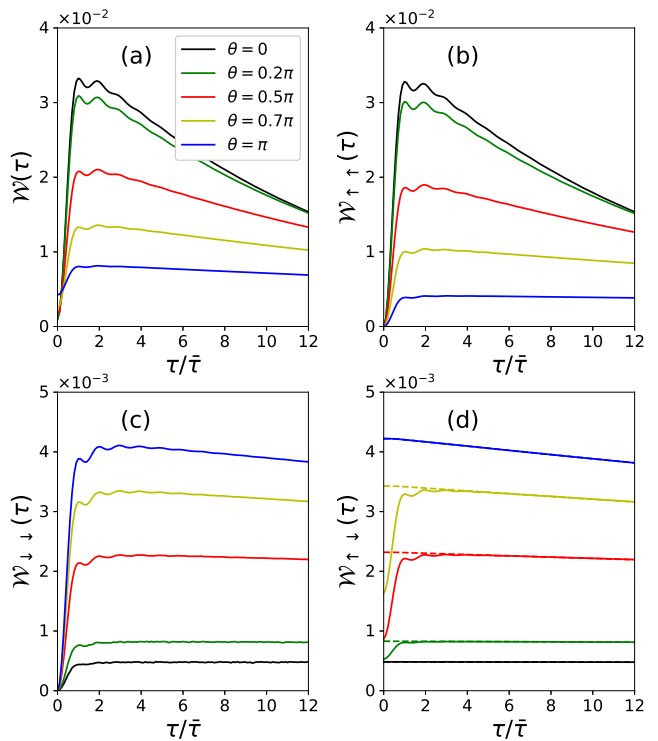} \\
  \caption{(Color online) WTDs for off-resonant transport corresponding to Fig.~\ref{fig3} with $\epsilon_\uparrow =\epsilon_\downarrow =5\Gamma$. }
  \label{fig5-off}
\end{figure}

	In this section, numerical outcome on the WTDs of the QD spin valve by varying noncollinear angle $\theta$ and spin polarization $p$ is reported. We choose the lead coupling strength $\Gamma$ as the energy unit. Voltage bias $eV=3\Gamma$ is applied on the left lead. The QD levels are set within the transport window with $\epsilon_\uparrow=2.0\Gamma$ and $\epsilon_\downarrow =1.5\Gamma$ for resonant transport, with the only exceptional case for Fig.~\ref{fig5-off}, wherein $\epsilon_\uparrow=\epsilon_\downarrow=5.0\Gamma$ for off-resonant transport. The waiting time $\tau$ is in units of the fundamental time scale $\bar{\tau}=h/(eV)$, which is the average time separation of the emitted electrons from the left lead.

	In Fig.~\ref{fig3}, we plot the two-channel WTD ${\cal W}(\tau)$ [panel (a)], spin-up WTD ${\cal W}_{\uparrow\uparrow}(\tau)$ [panel (b)], spin-down WTD ${\cal W}_{\downarrow\downarrow}(\tau)$ [panel (c)], and cross-channel WTD ${\cal W}_{\uparrow\downarrow}(\tau)$ [panel (d)] by varying noncollinear angle $\theta$ with spin polarization $p=0.8$. The FPTDs for spin down electrons ${\cal F}_{\downarrow}(\tau)$ are plotted with dashed lines in panel (d). $\theta=0$ and $\theta=\pi$ corresponds to parallel and antiparallel configuration, respectively. Differently from the single-channel case where the Pauli exclusion principle does not allow two electrons to occupy the same state, two electrons from different spin channels can be detected at the same time, so that two-channel WTD is nonzero at $\tau=0$. Two-channel WTD takes its maximal value at $\tau=\bar{\tau}$, and this is the same as that of spinless system \cite{WTD2012}. 
	With a positive polarization, spin-up and down states are majority and minority states in the left lead, respectively. The spin-up current in the right lead have contributions from both the spin-up and spin-down electrons in the left lead. Increasing $\theta$ from $0$ to $\pi$, the contribution to $I_{R\uparrow}$ from the spin-up (majority state) electrons in the left lead decreases, and the contribution from the spin-down (minority state) electrons increases. The combined effect leads to a decreasing spin-up current $I_{R\uparrow}$ in the right lead with increasing $\theta$. Due to a similar argument, one can explain that spin-down current $I_{R\downarrow}$ increases with increasing $\theta$. The particle current, as the sum of spin-up and spin-down current decreases with increasing $\theta$. These current behaviors with respect to noncollinear angle $\theta$ are shown in Fig.~\ref{fig4} (a-c). As can be observed from Fig.~\ref{fig3} (a-c), two-channel WTD ${\cal W}(\tau)$ and spin-up WTD ${\cal W}_{\uparrow\uparrow}(\tau)$ decrease with increasing $\theta$ at initial short times which are around before $\tau=5\bar{\tau}$, and spin-down WTD ${\cal W}_{\downarrow\downarrow}(\tau)$ and cross-channel WTD ${\cal W}_{\uparrow\downarrow}(\tau)$ increase with increasing $\theta$ at initial short times. Comparing the behaviors between currents and WTDs, one can observe that both particle current and two-channel WTD at initial short times decrease with increasing $\theta$, and $I_{R\sigma}$ and ${\cal W}_{\sigma'\sigma}(\tau)$ at initial short times share the same monotonicity.
	 The maximum point of ${\cal W}_{\downarrow\downarrow}(\tau)$ shifts towards shorter times with increasing angle $\theta$ from $0$ to $\pi$ and this indicates the increasing of the tunnel magnitude to the spin down state in the right lead as well.
	
	In Fig.~\ref{fig3} (d), FPTDs for spin down ${\cal F}_{\downarrow}(\tau)$ are plotted using dashed lines in comparison with the corresponding cross-channel WTD ${\cal W}_{\uparrow\downarrow}(\tau)$. We can observe that FPTD and WTD coincides with each other for the collinear configurations with $\theta=0$ and $\theta=\pi$, since the two spin channels are uncorrelated. Once the spin valve is in the non-collinear setup, cross-channel WTD deviates from its corresponding FPTD, and this indicates the occurring of spin torque transfer during transport. Cross-channel WTDs are less than FPTDs at initially short times and this indicates the suppression of subsequent detection due to the correlation between two spin channels and Pauli exclusion principle. One can observe that FPTD is a Monotonically decreasing function with respect to the time, while cross-channel WTD may not have this property.

	In order to better demonstrate the influence of the first detection on the subsequent detection result from the perspective of cross-channel WTD, the influence degree quantity is defined as the cumulative absolute difference between cross-channel WTDs and FPTDs with the expression,
\begin{equation}
{\rm IF} = \sum_\sigma \int_0^\infty |{\cal W}_{\bar{\sigma}\sigma}(\tau)-{\cal F}_\sigma (\tau)| d\tau .
\end{equation}
We plot the influence degree versus $\theta$ by varying spin polarization $p$ in Fig.~\ref{fig4} (d), and the influence degree versus spin polarization $p$ by varying $\theta$ in panel Fig.~\ref{fig7} (d). One can see that the influence degree vanishes for linear configurations with $\theta =0$ and $\theta=\pi$, and reaches its maximum near angle $\theta =\pi/2$ in which STT also achieves its maximal value \cite{Theodonis1,Theodonis2,Theodonis3}. 	

If one tunes the QD levels far away from the resonant condition, it can mimic the behaviors of MTJ with an insulating scattering region. Yu {\it et. al.} has shown that the off-resonant behaviors of the spin torque of QD spin valve is the same as that of MTJ \cite{YunjinYu}. WTDs for off-resonant transport corresponding to Fig.~\ref{fig3} with $\epsilon_\uparrow =\epsilon_\downarrow =5\Gamma$ are plotted in Fig.~\ref{fig5-off}. One can see that there are small oscillations with period $\bar{\tau}$ for all the WTDs shown in Fig.~\ref{fig5-off} at initial short times, due to the small transmission amplitude in the off-resonant condition. \cite{WTD2012} The behaviors of WTDs at short times with respect to noncollinear angle $\theta$.

\begin{figure}
\centering
  \includegraphics[width=3.4in]{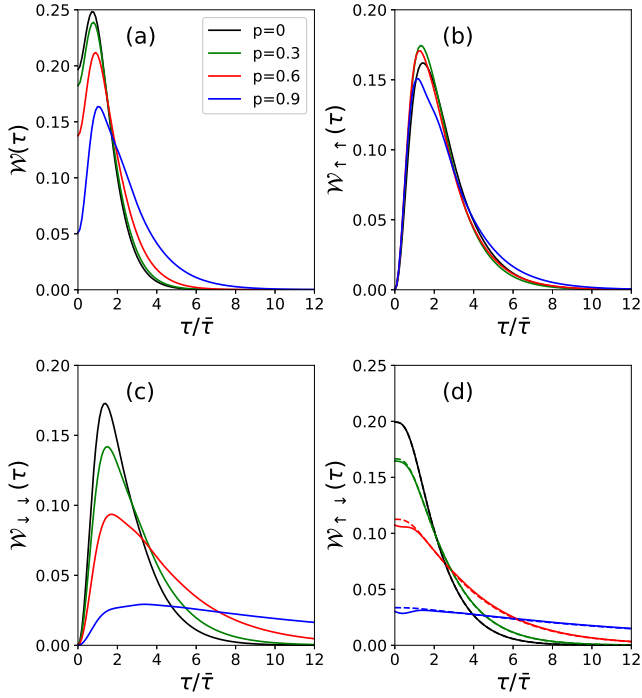} \\
  \caption{(Color online)  Two-channel WTD ${\cal W}(\tau)$ [panel (a)], spin-up WTD ${\cal W}_{\uparrow\uparrow}(\tau)$ [panel (b)], spin-down WTD ${\cal W}_{\downarrow\downarrow}(\tau)$ [panel (c)], and cross-channel WTD ${\cal W}_{\uparrow\downarrow}(\tau)$ [panel (d)] are plotted by varying spin polarization $p$ with $\theta=\pi/2$. 
The corresponding FPTD for spin down ${\cal F}_{\downarrow}(\tau)$ is plotted with dashed line in panel (d). }
  \label{fig6}
\end{figure}

\begin{figure}
  \includegraphics[width=3.5in]{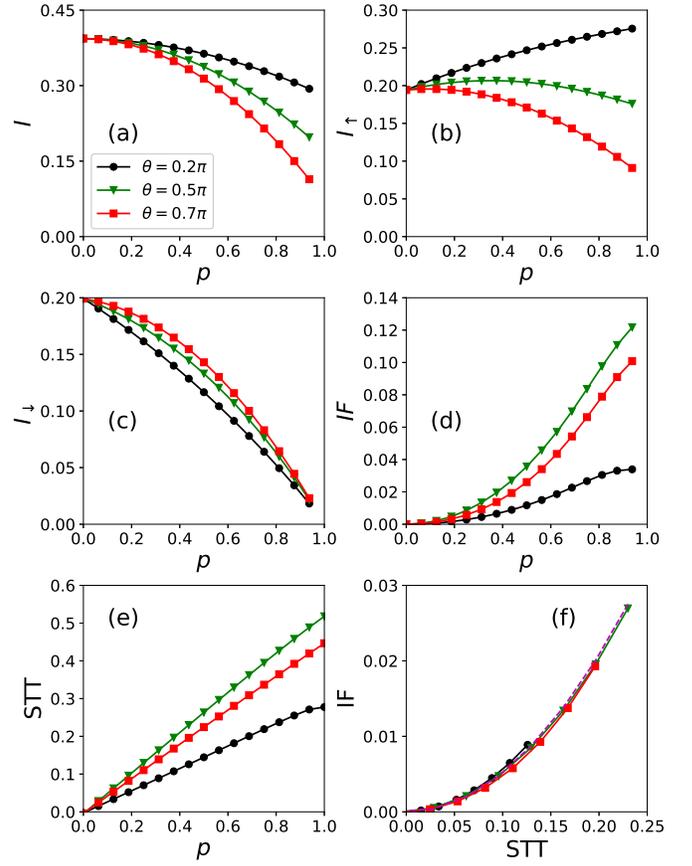} \\
  \caption{(Color online) Charge current [panel (a)], spin up current $I_{R\uparrow}$ [panel (b)], spin down current $I_{R\downarrow}$ [panel (c)], influence degree [panel (d)], and STT [panel (e)] versus spin polarization $p$ with different noncollinear angles $\theta$ are plotted. Influence degree versus STT are plotted in panel (f) with the quadratic fitting ${\rm IF}=0.51*({\rm STT})^2$ shown in dashed lines.  }
  \label{fig7}
\end{figure}

	In Fig.~\ref{fig6}, we plot the WTDs by varying spin polarization $p$ with noncollinear angle $\theta=\pi/2$ for the resonant transport. The corresponding FPTDs ${\cal F}_{\downarrow}(\tau)$ are plotted with dashed lines in panel (d). Increasing the polarization reduces both the particle current and spin-down electronic current so that two-channel WTD ${\cal W}(\tau)$, spin-down WTD ${\cal W}_{\downarrow\downarrow}(\tau)$ and cross-channel WTD ${\cal W}_{\uparrow\downarrow}(\tau)$ decrease at initial short times with increasing $p$. As indicated in Fig.~\ref{fig7} (b), the spin-up current is not monotonic with respect to polarization $p$ at $\theta=0.5\pi$, so is the short time behavior of spin-up WTD ${\cal W}_{\uparrow\uparrow}(\tau)$. The short time behavior of ${\cal W}_{\sigma'\sigma}(\tau)$ is the same as spin-$\sigma$ current $I_{R\sigma}$ by changing polarization $p$ by comparing Fig.~\ref{fig6} and Fig.~\ref{fig7}. 
	With $p\rightarrow 1$, both leads become half-metallic ferromagnet with only spin-up channel, ${\cal W}(\tau=0) \rightarrow 0$ due to Pauli exclusion principle. When both leads are normal metal, i.e., $p=0$, spin-up and -down channels are uncorrelated and there is no spin flip process through the junction, and as can be seen from Fig.~\ref{fig6} (d), cross-channel WTD ${\cal W}_{\uparrow\downarrow}(\tau)$ coincides with FPTD ${\cal F}_{\downarrow}(\tau)$. One can observe from  Fig.~\ref{fig7} (d) that the influence degree is zero for $p=0$, and this is independent of angle $\theta$. 
	Influence degree is an increment function with respect to the spin polarization regardless of noncollinear angle, as can be seen from panel Fig.~\ref{fig4} (d) and Fig.~\ref{fig7} (d). Cross-channel WTDs are less than FPTDs at initially short times by varying polarization, and this is also due to spin channel correlation.

	We plot the corresponding spin transfer torque (STT) versus noncollinear angle $\theta$ with different spin polarization $p$ in Fig. 4(e), and versus spin polarization with different noncollinear angle $\theta$ in Fig. 7(e). In order to show that spin-resolved waiting times can directly reflect STT, we show a one to one correspondence between STT and the influence degree quantity (IF) in panel in Fig.~\ref{fig4}(f) and Fig.~\ref{fig7}(f). The noncollinear angle $\theta$ takes the value in the range $[0,\pi/2]$ in Fig.~\ref{fig4}(f). 
	We can see that by varying noncollinear angle $\theta$ and polarization $p$, there is always a specific value of influence degree 'IF' corresponding to a given STT. We also find that they can be very well fitted by a quadratic relation ${\rm IF}=0.51*({\rm STT})^2$ of which the lines are plotted in dashed lines in Fig.~\ref{fig4}(f) and Fig.~\ref{fig4}(f). 
Finally, from Fig.~\ref{fig4}(f) and Fig.~\ref{fig4}(f) we see that IF versus STT for different noncollinear angles as well as different polarizations collapse into to a single curve showing universal behaviors. This demonstrates that the behaviors of spin-resolved waiting times are closely related STT.

\section{conclusion}\label{sec4}
In this work, we employ the scattering matrix approach to study the WTDs in a QD spin valve at zero temperature. 
The WTDs and FPTDs are calculated by taking derivatives with respect to the ITP which is a determinant involving both the spin and energy space.
The behaviors of two-channel WTD, spin-up WTD, spin-down WTD and cross-channel WTD are numerically shown with respect to noncollinear angle and spin polarization. 
Two-channel WTD takes its maximal value at $\tau=\bar{\tau}$, and is nonzero at $\tau=0$ which is due to the possibility of detecting two electrons from different spin channels at the same time. 
The short time behaviors of two-channel WTD and ${\cal W}_{\sigma'\sigma}(\tau)$ are the same as particle current and spin-$\sigma$ current $I_{R\sigma}$, respectively.	
	We observe that FPTD and WTD coincides with each other for the collinear configurations wherein the spin channels are uncorrelated. When the spin valve is in the non-collinear setup, the deviation of cross-channel WTD from its corresponding FPTD indicates the occurring of spin torque transfer across the junction. Cross-channel WTD is less than the corresponding FPTD at initially short times and this indicates the suppression of subsequent detection due to the correlation between the two spin channels. We introduce the 'influence degree' quantity to quantitatively characterize the correlation strength of spin channels. 
We find that the influence degree vanishes for linear configurations, and reaches its maximum near angle $\theta =\pi/2$ in which STT also achieves its maximal value. Since spin correlation strength increases with increasing spin polarization, influence degree is an increment function with respect to the spin polarization. We've shown that influence degree quantity is non-vanishing for the systems with spin-flip process.
	This work enables us to see that cross-channel WTD can be a new pathway to characterize properties in spintronics and motivates us to study waiting time distribution in other spintronics systems in the future.

\section{acknowledgments}
F. Xu is financially supported by the National Natural Science Foundation of China (Grants No. 11504240), and Shenzhen Key Lab Fund (20170228105421966). G. Tang and J. Wang are financially supported by NSF-China (Grant No. 11374246), the GRF (Grant No. 17311116), and the UGC (Contract No. AoE/P-04/08) of the Government of HKSAR.


\begin{thebibliography}{99}
\bibitem{spin1} S. A. Wolf, and D. Treger, IEEE Trans. Magn. {\bf 36}, 2748 (2000).
\bibitem{spin2} S. A. Wolf, D. D. Awschalom, R. A. Buhrman, J. M. Daughton, S. von Molna\'{a}r, M. L. Roukes, A. Y. Chtchelkanova, and D. M. Treger, Science {\bf 294}, 1488 (2001).
\bibitem{spin3} I. \v{Z}uti\'{c}, J. Fabian, and S. Das Sarma, Rev. Mod. Phys. {\bf 76}, 323 (2004).

\bibitem{MTJ1} J. K\"{o}nig, and J. Martinek, Phys. Rev. Lett. {\bf 90}, 166602 (2003).
\bibitem{MTJ2} M. Braun, J. K\"{o}nig, and J. Martinek, Phys. Rev. B {\bf 70}, 195345 (2004).
\bibitem{KGong} K. Gong, L. Zhang, L. Liu, Y. Zhu, G. Yu, P. Grutter, and H. Guo, J. Appl. Phys 118, 093902 (2016).

\bibitem{TMR1} M. Julliere, Phys. Lett. A {\bf 54}, 225 (1975).
\bibitem{TMR2} J.~C. Slonczewski, Phys. Rev. B {\bf 39}, 6995 (1989).
\bibitem{TMR3} S. S. P. Parkin {\it et al.}, Nat. Mater. {\bf 3}, 862 (2004).
\bibitem{TMR4} J. C. Sankey {\it et al.}, Nature Phys. {\bf 4}, 67 (2008).
\bibitem{TMR5} J. Mathon, and A. Umerski, Phys Rev. B, {\bf 63}, 220403(R) (2001).

\bibitem{STT1} J. C. Slonczewski, J. Magn. Magn. Mater. {\bf 159}, L1 (1996).
\bibitem{STT2} L. Berger, Phys. Rev. B {\bf 54}, 9353 (1996).

\bibitem{Theodonis1} I. Theodonis, N. Kioussis, A. Kalitsov, M. Chshiev, and W.~H.~Butler, Phys. Rev. Lett. {\bf 97}, 237205 (2006).
\bibitem{Theodonis2} I. Theodonis, A. Kalitsov, and N. Kioussis, Phys. Rev. B {\bf 76}, 224406 (2007).
\bibitem{Theodonis3} A. Kalitsov, M. Chshiev, I. Theodonis, N. Kioussis, and W.~H.~Butler, Phys. Rev. B {\bf 79}, 174416 (2009).

\bibitem{STT11} X. Jia, K. Xia, and G. E. W. Bauer, Phys. Rev. Lett. {\bf 107}, 176603 (2011).
\bibitem{STT22} H. Kubota {\it et al.}, Nature Phys. {\bf 4}, 37 (2008).

\bibitem{STT3} D. C. Ralph, and M. D. Stiles, J. Magn. Magn. Mater. {\bf 320}, 1190 (2008).

\bibitem{DiVentra} Y.~Dubi, and M.~D.~Ventra, Phys. Rev. B {\bf 79}, 081302(R) (2009).
\bibitem{Barnas} R.~\'{S}wirkowicz, M. Wierzbicki, and J.~Barna\'{s}, Phys. Rev. B {\bf 80}, 195409 (2009).
\bibitem{Bauer2} G. E. W. Bauer, E. Saitoh, and B. J. van Wees, Nature Mater. {\bf 11}, 391 (2012).
\bibitem{Bauer3} A. B. Cahaya, O. A. Tretiakov, and G. E. W. Bauer, IEEE Trans. Magn. {\bf 51}, 0800414 (2015).
\bibitem{gm5} G. Tang, J. Thingna, M. Esposito, and J. Wang, (in preparation). 

\bibitem{angle1} J. Fransson, Phys. Rev. B {\bf 72}, 045415 (2005).
\bibitem{angle2} J. N. Pedersen, J. Q. Thomassen, and K. Flensberg, Phys. Rev. B {\bf 72}, 045341 (2005).
\bibitem{angle3} I. Weymann, and J.~Barna\'{s}, Phys. Rev. B {\bf 75}, 155308 (2007).

\bibitem{pump} J. Splettstoesser, M. Governale, and K\"{o}nig, Phys. Rev. B {\bf 77}, 195320 (2008).
\bibitem{FCS-STT} P. Virtanen, and T. T. Heikkil\"{a}, Phys. Rev. Lett. {\bf 118}, 237701 (2017).

\bibitem{YunjinYu} Y. Yu, H. Zhan, L. Wan, B. Wang, Y. Wei, Q. Sun, and J. Wang, Nanotechnology {\bf 24}, 155202, (2013).

\bibitem{Blanter} Ya. Blanter, and M.~B\"{u}ttiker, Phys. Rep. {\bf 336}, 1 (2000).

\bibitem{Levitov1} L. S.~Levitov, and G. B.~Lesovik, Pis'ma Zh. Eksp. Teor. Fiz. {\bf 58}, 225 (1993) [Sov. Phys. JETP Lett. {\bf 58}, 230 (1993)].
\bibitem{Levitov2} L. S.~Levitov, H.-W.~Lee, and G. B.~Lesovik, J. Math. Phys. {\bf 37}, 4845 (1996).
\bibitem{Flindt1} C Flindt, T. Novotn\'{y}, A. Braggio, M. Sassetti, and A. P. Jauho, Phys. Rev. Lett. {\bf 100}, 150601 (2008).
\bibitem{wavepacket} F.~Hassler, M. V.~Suslov, G. M.~Graf, M. V.~Lebedev, G. B.~Lesovik, and G.~Blatter, Phys. Rev. B {\bf 78}, 165330 (2008).
\bibitem{Flindt2} C. Flindt, T.~Novotn\'{y}, A. Braggio, and A. P. Jauho, Phys. Rev. B {\bf 82}, 155407 (2010).
\bibitem{Fernando} F. Dom\'{i}nguez, G. Platero, S. Kohler, J. Chem. Phys. {\bf 375}, 284 (2010).
\bibitem{Flindt3} V. F. Maisi, D. Kambly, C. Flindt, and J. P. Pekola, Phys. Rev. Lett. {\bf 112}, 036801 (2014).
\bibitem{gm2} G.-M.~Tang, and J.~Wang, Phys. Rev. B {\bf 90}, 195422 (2014).
\bibitem{JS3} B. K.~Agarwalla, H.~Li, B.~Li, and J.-S.~Wang, Phys. Rev. E {\bf 89}, 052101 (2014).
\bibitem{Ruben1} R.~S.~Souto, A.~Mart\'{i}n-Rodero, and A.~L.~Yeyati, Phys. Rev. Lett. {\bf 117}, 267701 (2016).
\bibitem{gm3} G. Tang, Z. Yu, and J. Wang, New J. Phys. {\bf 19}, 083007 (2017).
\bibitem{gm4} G.~Tang, Y.~Xing, and J.~Wang, Phys. Rev. B {\bf 96}, 075417 (2017).
\bibitem{gm6} G. Tang, X. Chen, J. Ren, and J. Wang, arXiv:1705.10025.
\bibitem{Ruben2} R.~S.~Souto, A.~Mart\'{i}n-Rodero, and A.~L.~Yeyati, Phys. Rev. B {\bf 96}, 165444 (2017).
\bibitem{Michael} M. Ridley, V. N. Singh, E. Gull, and G. Cohen, arXiv:1801.05010.

\bibitem{FCS-MTJ} S. Lindebaum, D. Urban, and J. K\"{o}nig, Phys. Rev. B {\bf 79}, 245303 (2009).
\bibitem{switch} G. Utsumi, and T. Taniguchi, Phys. Rev. Lett. {\bf 114}, 186601 (2015).

\bibitem{SET} M. A. Kastner, Rev. Mod. Phys, {\bf 64}, 3 (1992); D. L. Klein {\it et al.}, Nature (London) {\bf 389}, 699 (1997); J. Martin {\it et al.}, Nature Phys. {\bf 4}, 144 (2008).


\bibitem{kampen}
N. G.~van Kampen, \emph{Stochastic Processes in Physics and Chemistry} (Elsevier, Amsterdam, 2007).
\bibitem{WTD_Brandes}
T. Brandes, Ann. Phys. {\bf 17}, 477 (2008).

\bibitem{WTD_Yan}
S. Welack, S. Mukamel, and Y. Yan, Europhys. Lett. {\bf 85}, 57008 (2009).

\bibitem{WTD2011}
M.~Albert, C.~Flindt, and M.~B\"{u}ttiker, Phys. Rev. Lett. {\bf 107}, 086805 (2011).

\bibitem{WTD_rajabi}
L. Rajabi, C. Poltl, and M. Governale, Phys. Rev. Lett. {\bf 111}, 067002, (2013).

\bibitem{WTD_spin_valve} B. Sothmann, Phys. Rev. B {\bf 90}, 155315 (2014).

\bibitem{WTD_Kosov}
D. S. Kosov, J. Chem. Phys. {\bf 146}, 074102 (2017).

\bibitem{WTD_turnstile} 
E. Potanina, and C. Flindt, Phys. Rev. B {\bf 96}, 045420 (2017).

\bibitem{WTD_CPS}
N. Walldorf, C. Padurariu, A.-P. Jauho, and C. Flindt, arXiv:1709.01335.

\bibitem{WTD_non-Markovian} K. H. Thomas, and C. Flindt, Phys. Rev. B {\bf 87}, 121405 (2013).


\bibitem{WTD2012}
M.~Albert, G.~Haack, C.~Flindt, and M.~B\"{u}ttiker, Phys. Rev. Lett. {\bf 108}, 186806 (2012).

\bibitem{WTD_TB}
K. H. Thomas, and C. Flindt, Phys. Rev. B {\bf 89}, 245420 (2014).

\bibitem{WTD_Floquet}
D. Dasenbrook, C. Flindt, and M. Buttiker, Phys. Rev. Lett. {\bf 112}, 146801 (2014).

\bibitem{WTD_Leviton}
M. Albert, and P. Devillard, Phys. Rev. B {\bf 90}, 035431 (2014).

\bibitem{WTD2}
P. P. Hofer, D. Dasenbrook, and C. Flindt, Physica E {\bf 82}, 3 (2016).


%

\bibitem{WTD_clock}
D. Dasenbrook, and C. Flindt, Phys. Rev. B {\bf 93}, 245409 (2016).


\bibitem{WTD2014}
G. Haack, M. Albert, and C. Flindt, Phys. Rev. B {\bf 90}, 205429 (2014).

\bibitem{WTD_correlated}
D. Dasenbrook, P. P. Hofer, and C. Flindt, Phys. Rev. B {\bf 91}, 195420 (2015).

\bibitem{QSH_QD1} C. Timm, Phys. Rev. B, {\bf 86}, 155456, (2012).
\bibitem{QSH_QD2} G. Dolcetto, N. T. Ziani, M. Biggio, F. Cavaliere, and M. Sassetti, Phys. Rev. B, {\bf 87}, 235423, (2012).
\bibitem{QSH_QD3} N. T. Ziani, C. Fleckenstein, G. Dolcetto, and B. Trauzettel, Phys. Rev. B, {\bf 95}, 205418, (2017).

\bibitem{Datta} S. Datta, {\it Electronic Transport in Mesoscopic Systems} (Cambridge University Press, Cambridge, 1997).

\bibitem{JianWang} J. Wang, and H. Guo, Phys. Rev. B {\bf 79}, 045119 (2009).

\bibitem{off-diagonal} A. Nock, S. Kumar, H.-J. Sommers, and T. Guhr, Ann. Phys. {\bf 342}, 103 (2014). 

\bibitem{gm1} G.-M.~Tang, F.~Xu, and J.~Wang, Phys. Rev. B {\bf 89}, 205310 (2014).


\bibitem{double-QWs} P. Michetti, J. C. Budich, E. G. Novik, and P. Recher, Phys. Rev. B, {\bf 85}, 125309, (2012).

\bibitem{QSH_QD4}  B. Rizzo, A. Camjayi, and L. Arrachea, Phys. Rev. B {\bf 94}, 125425 (2016).
\bibitem{QSH_QD5} Y. Xing, Z.-l. Yang, Q.-f. Sun, and J. Wang, Phys Rev. B {\bf 90}, 075435 (2014).

\end{thebibliography}
\end{document}